\documentclass[letterpaper,11pt]{article}

\usepackage[latin1]{inputenc}
\usepackage[T1]{fontenc}
\usepackage{graphicx}
\usepackage{amsfonts}
\usepackage{amssymb}
\usepackage{amsmath} 
\usepackage{amsthm}
\usepackage{url}

\usepackage[margin=1in]{geometry}

\newcommand {\br}[1]{\left(#1\right)}
\newcommand {\cbr}[1]{\left\{#1 \right\}}
\newcommand{\RR}{\mathbb{R}}    
\newcommand{\Scal}{\mathcal{S}} 
\newcommand{\Tcal}{\mathcal{T}} 
\newcommand {\sqb}[1]{\left[#1\right]}

\date{June 1, 2008\\Revised September 22, 2008}

\begin{document}

\title{Reconstruction of biological networks by supervised machine learning approaches}
\author{
Jean-Philippe Vert$^{1,2,3}$\\
$^1$ Mines ParisTech, Centre for Computational Biology\\
35, rue Saint-Honor\'e\\
F-77300 Fontainebleau, France;\\
$^2$ Institut Curie, Paris, F-75248;\\
$^3$ INSERM, U900, Paris, F-75248.\\
\texttt{jean-philippe.vert@ensmp.fr}
}

\maketitle

\begin{abstract}
We review a recent trend in computational systems biology which aims at using pattern recognition algorithms to infer the structure of large-scale biological networks from heterogeneous genomic data. We present several strategies that have been proposed and that lead to different pattern recognition problems and algorithms. The strength of these approaches is illustrated on the reconstruction of metabolic, protein-protein and regulatory networks of model organisms. In all cases, state-of-the-art performance is reported.
\end{abstract}

\section[Introduction]{Introduction}

In this review chapter we focus on the problem of reconstructing the structure of large-scale biological networks. By biological networks we mean graphs whose vertices are all or a subset of the genes and proteins encoded in a given organism of interest, and whose edges, either directed or undirected, represent various biological properties. As running examples we consider the three following graphs, although the methods presented below may be applied to other biological networks as well.
\begin{itemize}
\item \emph{Protein-protein interaction (PPI) network}. This is an undirected graph with no self-loop, that contains all proteins encoded by an organism as vertices. Two proteins are connected by an edge if they can physically interact.
\item \emph{Gene regulatory network}. This is a directed graph that contains all genes of an organism as vertices. Among the genes, some called transcription factors (TFs) regulate the expression of other genes through binding to the DNA. The edges of the graph connect TFs to the genes they regulate. Self-loops are possible if a TF regulates itself. Moreover each edge may in principle be labeled to indicate whether the regulation is a positive (activation) or negative (inhibition) regulation.
\item \emph{Metabolic network}. This graph contains only a subset of the genes as vertices, namely those coding for enzymes. Enzymes are proteins whose main function is to catalyse a chemical reaction, transforming substrate molecules into product molecules. Two enzymes are connected in this graph if they can catalyse two successive reactions in a metabolic pathway, i.e., two reactions such that the main product of the first one is a substrate of the second one.
\end{itemize}
Deciphering these networks for model organisms, pathogens or human is currently a major challenge in systems biology, with many expected applications ranging from basic biology to medical applications. For example, knowing the detailed interactions possible between proteins on a genomic scale would highlight key proteins that interact with many partners, which could be interesting drug targets \cite{Jeong2001Lethality}, and would help in the annotation of proteins by annotation transfer between interacting proteins. The elucidation of gene regulatory networks, especially in bacteria and simple eukaryotes, would provide new insights into the complex mechanisms that allow an organism to regulate its metabolism and adapt itself to environmental changes, and could provide interesting guidelines for the design of new functions.  Finally, understanding in detail the metabolism of an organism, and clarifying which proteins are in charge of its control, would give a valuable description of how organisms have found original pathways for degradation and synthesis of various molecules, and could help again in the identification of new drug targets \cite{Okamoto2007Prediction}.

Decades of research in molecular biology and genetics have already provided a partial view of these networks, in particular for model organisms. Moreover, recent high-throughput technologies such as the yeast two-hybrid systems for PPI, provide large numbers of likely edges in these graphs, although probably with a high rate of false positives \cite{Uetz2000comprehensive,Ito2001comprehensive}. Thus, much work remains to be done in order to \emph{complete} (adding currently unknown edges) and \emph{correct} (removing false positive edges) these partially known networks. To do so, one may want to use information about individual genes and proteins, such as their sequence, structure, subcellular localization, or level of expression across several experiments. Indeed, this information often provides useful hints about the presence or absence of edges between two proteins. For example, two proteins are more likely to interact physically if they are expressed in similar experiments, and localized in the same cellular compartment; or two enzymes are more likely to be involved in the same metabolic pathway if they are often co-expressed, and if they have homologs in the same species \cite{Marcotte1999Detecting,Pazos2001Similarity,Jansen2003Bayesian}.

Following this line of thought, many approaches have been proposed in the recent years to infer biological networks from genomic and proteomic data, most of them attempting to reconstruct the graphs \emph{de novo}. In \emph{de novo} inference, the data about individual genes and proteins are given, and edges are inferred from these data only, using a variety of inference principles. For example, when time series of expression data are used, regulatory networks have been reconstructed by fitting various dynamical system equations to the data \cite{Akutsu2000Algorithms,Chen1999Modeling,Tegner2003Reverse,Gardner2003Inferring,Chen2005stochastic,Bernardo2005Chemogenomic,Bansal2006Inference}. Bayesian networks have also been used to infer \emph{de novo} regulatory networks from expression data, assuming that direct regulation can be inferred from the analysis of correlation and conditional independence between expression levels \cite{Friedman2000Using}. Another rationale for \emph{de novo} inference is to connect genes or proteins that are \emph{similar} to each other in some sense \cite{Marcotte1999Detecting,Pazos2001Similarity}, For example, co-expression networks, or the detection of similar phylogenetic profiles are popular ways to infer "functional relationships" between proteins, although the meaning of the resulting edges has no clear biological justification \cite{Tavazoie1999Systematic}. Similarly, some authors have attempted to predict gene regulatory networks by detecting large mutual information between expression levels of a TF and the genes it regulates \cite{Butte2000Discovering,Faith2007Large-scale}.

In contrast to these \emph{de novo} methods, in this review we present a general approach to reconstruct biological networks using information about individual genes and proteins, based on \emph{supervised} machine learning algorithms, as developed through a recent series of articles \cite{Yamanishi2004Protein,Vert2005Supervised,Yamanishi2005Supervised,Ben-Hur2005Kernel,Biau2006Statistical,Vert2007new,Bleakley2007Supervised,Mordelet2008SIRENE}. The graph inference paradigm we follow assumes that, besides the information about individual vertices (genes or proteins) used by \emph{de novo} approaches, the graph we wish to infer is also partially known, and known edges can be used by the inference algorithm to infer unknown edges. This paradigm is similar to the notion of \emph{supervised inference} in statistics and machine learning, where one uses a set of input/output pairs (often called the training set) to estimate a function that can predict the output associated to new inputs \cite{Hastie2001elements,Bishop2006Pattern}. In our paradigm, we give us the right to use the known edges of the graph to supervise the estimation of a function that could predict whether a new pair of vertices is connected by an edge or not, given the data about the vertices.  Intuitively, this setting can allow us to automatically learn what features of the data about vertices are the most informative to predict the presence of an edge between two vertices. In a sense, this paradigm leads to a problem much simpler than the \emph{de novo} inference problem, since more information is used as input, and it might seem unfair to compare \emph{de novo} and supervised methods. However, as already mentioned, in many real-world cases of interest we already partially know the graph we wish to infer. It is therefore quite natural to use as much information as we can in order to focus on the real problem, which is to infer new edges (and perhaps delete wrong edges), and therefore to use as input both the genomic and proteomic data, on the one hand, and the edges already known, on the other hand.

In a slightly more formal language, we therefore wish to learn a function that can predict whether an edge exists or not between two vertices (genes or proteins), given data about the vertices (e.g., expression levels of each gene in different experimental conditions). Technically this problem can be thought of as a problem of \emph{binary classification}, where we need to assign a binary label (presence or absence of edge) to each pair of vertices, as explained in Section \ref{sec:problem}.  From a computational point of view, the supervised inference paradigm we investigate can in principle benefit from the availability of a number of methods for supervised binary classification, also known as pattern recognition \cite{Bishop2006Pattern}. These methods, as reviewed in Section \ref{sec:pattrec} below, are able to estimate a function to predict a binary label from data about patterns, given a training set of (pattern, label) pairs. The supervised inference problem we are confronted with, however, is not a classical pattern/label problem, because the data are associated to individual vertices (e.g., expression profiles are available for each individual gene), while the labels correspond to pairs of vertices. Before applying out of the box state-of-the-art machine learning algorithms, we therefore need to clarify how our problem can be transformed as a classical pattern recognition problem (Section \ref{sec:graphpatrec}). In particular, we show that there is not a unique way to do that and present in Sections \ref{sec:local} and \ref{sec:global} two classes of approaches that have been proposed recently. Both classes involve a support vector machine (SVM) as binary classification engine, but follow different avenues to cast the edge inference problem as a binary classification problem. In Section \ref{sec:exp}, we provide experimental results that justify the relevance of supervised inference, and show that a particular approach, based on local models, performs particularly well on the reconstruction of PPI, regulatory and metabolic networks. We conclude with a rapid discussion in Section \ref{sec:discussion}.

\section{Graph reconstruction as a pattern recognition problem}\label{sec:pat}

In this section we define formally the graph reconstruction problem considered, and explain how to solve it with pattern recognition techniques.

\subsection{Problem formalization}\label{sec:problem}
We consider a finite set of vertices $V=\br{v_{1},\ldots,v_{n}}$ that typically correspond to the set of all genes or proteins of an organism. We further assume that for each vertex $v\in V$ we have a description of various features of $v$ as a vector $\phi(v)\in\RR^p$. Typically, $\phi(v)$ could be a vector of expression levels of the gene $v$ in $p$ different experimental conditions, measured by DNA microarrays, a phylogenetic profile which encodes the presence or absence of the gene in a set of $p$ sequenced genomes \cite{Pazos2001Similarity}, a vector of $p$ sequence features, or a combination of such features. We wish to reconstruct a set of edges $E\subset V\times V$ that defines a biological network. While in \emph{de novo} inference the goal is to design an algorithm that automatically predicts edges in $E$ from the set of vertex features $\br{\phi(v_{1}),\ldots,\phi(v_{n})}$, in our approach we further assume that a set of pairs of vertices known to be connected by an edge or not is given. In other words we assume given a list $\Scal=\br{(e_{1},y_{1}),\ldots,(e_{N},y_{N})}$ of pairs of vertices ($e_{i}\in V\times V$) tagged with a label $y_{i}\in\cbr{-1,1}$ that indicate whether the pair $e_{i}$ is known to interact ($y_{i}=1$) or not ($y_{i}=-1$). In an ideal noise-free situation, where the labels of pairs in the training set are known with certainty, we thus have $y_{i}=1$ if $e_{i} \in E$,  and $y_{i}=-1$ otherwise. However, in some situations we may also have noise or errors in the training set labels, in which case we could only assume that pairs in $E$ tend to have a positive label, while pairs not in $E$ tend to have a negative label.

\begin{figure}[ht]
\begin{center}
\includegraphics[width=.5\textwidth]{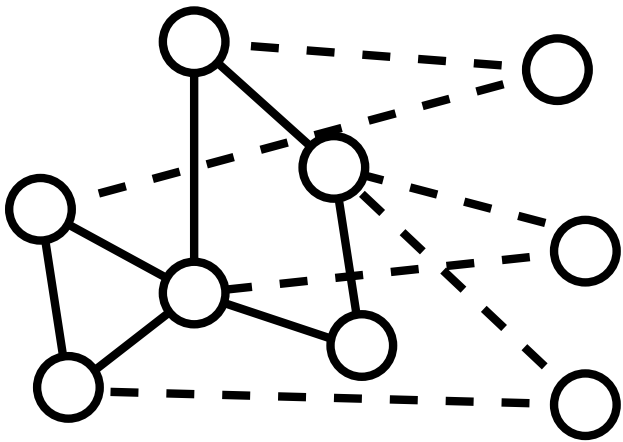}
\caption{We consider the problem of inferring missing edges in a graph (dotted edges) where a few edges are already known (solid edges). To carry out the inference, we use attributes available about individual vertices, such as vectors of expression levels across different experiments if vertices are genes.}
\label{fig:theproblem}
\end{center}
\end{figure}

The graph reconstruction problem can now be formally stated as follows: given the training set $\Scal$ and the set of vertex features $\br{\phi(v_{1}),\ldots,\phi(v_{n})}$, predict for all pairs not in $\Scal$ whether they interact (i.e., whether they are in $E$) or not. This formulation is illustrated in Figure \ref{fig:theproblem}.

Stated this way, this problem is similar to a classical pattern recognition problems, for which a variety of efficient algorithms have been developed over the years. Before highlighting the slight difference between the classical pattern recognition framework and ours, it is therefore worth recalling this classical pattern recognition paradigm and mentioning some algorithms adapted to solve it.

\subsection{Pattern recognition}\label{sec:pattrec}
Pattern recognition, of binary supervised classification, is a well-studied problem in statistics and machine learning \cite{Hastie2001elements,Bishop2006Pattern}. In its basic set-up, a training set $\Tcal=\cbr{(u_{1},t_{1}),\ldots,(u_{N},t_{N})}$ of labeled patterns is given,  where $u_{i} \in \RR^q$ is a vector and $t_{i}\in\cbr{-1,1}$ is a binary label, for $i=1,\ldots,N$. The goal is then to infer a function $f:\RR^q\rightarrow\cbr{-1,1}$ that is able to predict the binary label $t$ of any new pattern $u\in\RR^q$ by $f(u)$.

Many methods have been proposed to infer the labeling function $f$ from the training set $\Tcal$, including for example nearest neighbor classifiers, decision trees, logistic regression, artificial neural networks or support vector machines (SVM). Although any of these methods can be used in what follows, we will present experiments carried out with an SVM, which we briefly describe below, mainly for three reasons:
\begin{itemize}
\item It is now a widely-used algorithm, in particular in computational biology, with many public implementations \cite{Schoelkopf2004Kernel,Vert2007Kernel}.
\item It provides a convenient framework to combine heterogeneous features about the vertices, such as the sequence, expression and subcellular localization of proteins \cite{Pavlidis2002Learning,Yamanishi2004Protein,Lanckriet2004statistical}.
\item Some methods developed so far for graph inference, which we describe below, are particularly well adapted for a formalization in the context of SVM and kernel methods \cite{Ben-Hur2005Kernel,Vert2007new}.
\end{itemize}
Let us therefore briefly describe the SVM algorithm, and redirect the interested reader to various textbooks for more details \cite{Vapnik1998Statistical,Cristianini2000introduction,Scholkopf2002Learning}. Given the labeled training set $\Tcal$, an SVM estimates a linear function $h(u) = w^\top u$ for some vector $w\in\RR^q$ (here $w^\top u$ represents the inner product between $w$ and $u$), and then makes a label prediction for a new pattern $u$ that depends only on the sign of $h(u)$: $f(u) = 1$ if $h(u)\geq 0$, $f(u) = -1$ otherwise. The vector $w$ is obtained as the solution of an optimization problem that attempts to enforce a correct sign with large absolute values for the values $h(u_{i})$ on the training set, while controlling the Euclidean norm of $w$. The resulting optimization problem is a quadratic program for which many specific and fast implementations have been proposed. 

An interesting property of SVM, particularly for the purpose of heterogeneous data integration, is that the optimization problem only involves the training patterns $u_{i}$ through pairwise inner products of the form $u_{i}^\top u_{j}$. Moreover, once the classifier is trained, the computation of $h(u)$ to predict the label of a new point $u$ also involves only patterns through inner products of the form $u^\top u_{i}$. Hence, rather than computing and storing each individual pattern as a vector $u$, we just need to be able to compute inner products of the form $u^\top u'$ for any two patterns $u$ and $u'$ in order to train an SVM and use it as a prediction engine. This inner product between patterns $u$ and $u'$ is a particular case of what is called a \emph{kernel} and denoted $K(u,u') = u^\top u'$, to emphasize the fact that it can be seen as a function that associate a number to any pair of patterns $(u,u')$, namely their inner product. More generally a kernel is a function that computes the inner product between two patterns $u$ and $u'$ after possibly mapping them to some vector space with inner product by a mapping $\phi$, i.e., $K(u,u') = \phi(u)^\top \phi(u)'$.

Kernels are particularly relevant when the patterns are represented by vectors of large dimensions, whose inner products can nevertheless be computed efficiently. They are also powerful tools to integrate heterogeneous data. Suppose for example that each pattern $u$ can be represented as two different vectors $u^{(1)}$ and $u^{(2)}$. This could be the case, for example, if one wanted to represent a protein $u$ either by a vector of expression profile $u^{(1)}$ or by a vector of phylogenetic profile $u^{(2)}$. Let now $K_{1}$ and $K_{2}$ be the two kernels corresponding to inner products for each representation, namely, $K_{1}(u,u') = u^{(1)\top} u^{(1)'}$ and $K_{2}(u,u') = u^{(2)\top} u^{(2)'}$. If we now want to represent both types of features into a single representation, a natural approach would be, e.g., to \emph{concatenate} both vectors $u^{(1)}$ and $u^{(2)}$ into a single vector, which we denote by $u^{(1)} \oplus u^{(2)}$ (also called the \emph{direct sum} of $u^{(1)}$ and $u^{(2)}$). In order to use this joint representation in an SVM, we need to be able to compute the inner products between direct sums of two patterns to define a joint kernel $K_{joint}$. Interestingly, some simple algebra shows that the resulting inner product is easily expressed as the sum of the inner products of each representation, i.e.: 
\begin{equation}\label{eq:kernelsum}
\begin{split}
K_{joint}(u,u') & = \br{u^{(1)} \oplus u^{(2)}}^\top \br{u^{(1)'} \oplus u^{(2)'}} \\
& = \left(\begin{array}{c}u^{(1)} \\u^{(2)}\end{array}\right)^\top \left(\begin{array}{c}u^{(1)'} \\u^{(2)'}\end{array}\right) \\
&= u^{(1)^\top} u^{(1)'} + u^{(2)^\top} u^{(2)'} \\
&= K_{1}(u,u') + K_{2}(u,u')\,.
\end{split}
\end{equation}
Consequently, the painstaking operation of concatenation between two vectors of potentially large dimension is advantageously replaced by simply doing the sum between two kernels. More generally, if $k$ different representations are given, corresponding to $k$ different kernels, then summing together the $k$ kernels results in a joint kernel that integrates all different representations. The sum can also be replaced by any convex combination (linear combination with nonnegative weights) in order to weight differently the importance of different features \cite{Lanckriet2004statistical}.

\subsection{Graph inference as a pattern recognition problem}\label{sec:graphpatrec}
Let us now return to the graph reconstruction problem, as presented in Section \ref{sec:problem}. At first sight, this problem is very similar to the general pattern recognition paradigm recalled in Section \ref{sec:pattrec}: given pairs of vertices with positive and negative labels, infer a function $f$ to predict whether a new pair has a positive label (i.e., is connected) or not. An important difference between the two problems, however, is that the features available in the graph reconstruction problem describe properties of individual vertices $v$, and not of pairs of vertices $(v,v')$. Thus, in order to apply pattern recognition techniques such as the SVM  to solve the graph reconstruction problem, we can follow one of two possible avenues:
\begin{enumerate}
\item Reformulate the graph reconstruction problem as a pattern recognition problem where binary labels are attached to individual vertices (and not to pairs of vertices). Then pattern recognition methods can be used to infer the label of vertices based on their features.
\item Keep the formulation as the problem of predicting the binary label of a pair of vertices, but find a way to represent as vectors (or as a kernel) pairs of vertices, while we initially only have features for individual vertices.
\end{enumerate}
Both directions are possible and have been investigated by different authors, leading to different algorithms. In Section \ref{sec:local} we present an instantiation of the first idea, which rephrases graph reconstruction as a combination of simple pattern recognition problems at the level of individual vertices. In Section \ref{sec:global} we present several instantiations of the second strategies, which amount to defining a kernel for pairs of vertices from a kernel for individual vertices.

\subsection{Graph inference with local models}\label{sec:local}
\begin{figure}[ht]
\begin{center}
\includegraphics[width=.5\textwidth]{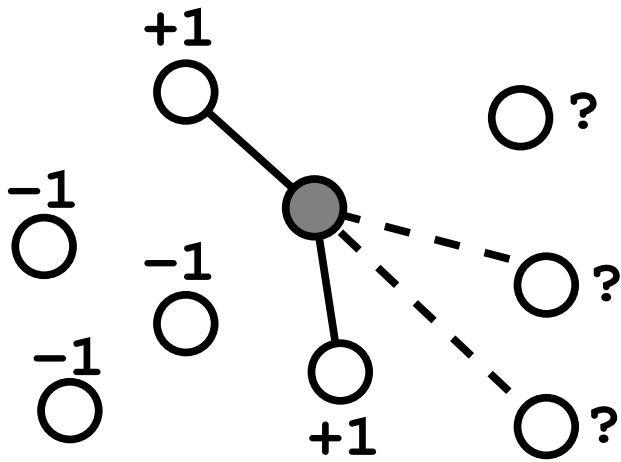}
\caption{Illustration of one binary classification problem that is generated from the graph inference problem of Figure \ref{fig:theproblem} with the local model approach. Taking the shaded vertex as seed, other vertices in the training set are labeled as $+1$ of $-1$ depending on whether they are known to be connected or to be not connected to the shaded vertex. The goal is then to predict the label of vertices not used during training. The process is then repeated by shading each vertex in turn.}
\label{fig:theproblem2}
\end{center}
\end{figure}

In this section we describe an approach that was proposed by \cite{Bleakley2007Supervised} for the reconstruction of metabolic and PPI networks, and also successfully applied by \cite{Mordelet2008SIRENE} for regulatory network inference. The basic idea is very simple, an can be thought of as a ``divide-and-conquer`` strategy to infer new edges in a graph. Each vertex of the graph is considered in turn as a seed vertex, independently from the others, and a ``local'' pattern recognition problem is solved to discriminate the vertices that are connected to this seed vertex against the vertices that are not connected to it. The local model can then be applied to predict new edges between the seed vertex and other vertices. This process is then repeated with other vertices as seed to obtain edge prediction throughout the graph. More precisely, the ``local model'' approach can be described as follows:
\begin{enumerate}
\item Take a seed vertex $v_{seed}$ in $V$.
\item For each pair $(v_{seed},v')$ with label $y$ in the training set, associate the same label $y$ to the individual vertex $v'$. This results in a set of labeled vertices $\cbr{(v'_{1},t_{1}),\ldots,(v'_{n(v_{seed})},t_{n(v_{seed})})}$, where $n(v_{seed})$ is the number of pairs starting with $v_{seed}$ in the training set. We call this set a local training set.
\item Train a pattern recognition algorithm on the local training set designed in step 2.
\item Predict the label of any vertex $v'$ that has no label, i.e., such that $(v_{seed},v')$ is not in the training set.
\item If a vertex $v'$ has a positive predicted label, then predict that the pair $(v_{seed},v')$ has a positive label (i.e., is an edge).
\item Repeat step 1-5 for each vertex $v_{seed}$ in $V$.
\item Combine the edges predicted at each iteration together, to obtain the final list of predicted edges.
\end{enumerate}
This process is illustrated in Figure \ref{fig:theproblem2}.
Intuitively, such an approach can work if the features about individual vertices provide useful information about whether or not they share a common neighbor. For example, the approach was developed by \cite{Mordelet2008SIRENE} to reconstruct the gene regulatory network, i.e., to predict whether a transcription factor $v$ regulates a gene $v'$, using a compendium of gene expression levels across a variety of experimental conditions as features. The paradigm seems particularly relevant in that case. Indeed, if two genes are regulated by the same TF, then they are likely to behave similarly in terms of expression level; conversely, if a gene $v'$ is known to be regulated by a TF $v$, and if the expression profile of another gene $v''$ is similar to that of $v'$, then one can predict that $v''$ is likely to be regulated by $v$. The pattern recognition algorithm is precisely the tool that automatizes the task of predicting that $v''$ has positive label, given that $v'$ has itself a positive label and that $v'$ and $v''$ share similar features.

We note that this local model approach is particularly relevant for directed graphs, such as gene regulatory networks. If our goal is to reconstruct an undirected graph, such the PPI graph, then one can follow exactly the same approach, except that (i) each undirected training pair $\{v,v'\}$ should be considered twice in step 2, namely as the directed pair $(v,v')$ for the local model of $v$ and as the directed pair $(v',v)$ for the directed model of $v'$, and (ii) in the prediction step for an undirected pair $\{v,v'\}$, the prediction of the label of the directed pair $(v,v')$ with the local model of $v$ must be combined with the prediction of the label of the directed pair $(v',v)$ made by the local model of $v'$. In \cite{Bleakley2007Supervised}, for example, in the prediction step the score of the directed pair $(v,v')$ is averaged with the score of the directed pair $(v',v)$ to obtain a unique score for the undirected pair $\{v,v'\}$.

In terms of computational complexity, it can be very beneficial to split a large pattern recognition problem into several smaller problems. Indeed, the time and memory complexities of pattern recognition algorithms such as SVM are roughly quadratic or worse in the number of training examples. If a training set of $N$ pairs is split into $s$ local training sets of roughly $N/s$ patterns each, then the total cost of running $s$ SVM to estimate local models will therefore be of the order of $s\times \br{N/s}^2 = N^2/s$. Hence if a local model is built for each vertex ($s=n$), one can expect a speed-up of the algorithm of up to a factor of $n$ over an SVM that would work with $N$ pairs as training patterns. Moreover, the local problems associated to different seed vertices being independent from each others, one can trivially benefit from parallel computing architectures by training the different local models on different processors. 

On the other hand, an apparently important drawback of the approach is that the size of each local training set can become very small if, for example, a vertex has few or even no known neighbors. Inferring accurate predictive models from few training examples is known to be challenging in machine learning, and in the extreme case where a vertex has no known neighbor during training, then no new edge can ever be predicted. However, the experimental results, reported by \cite{Bleakley2007Supervised,Mordelet2008SIRENE} and in Section \ref{sec:exp}, show that one can obtain very competitive results with local models in spite of this apparent difficulty.

\subsection{Graph inference with global models}\label{sec:global}
Splitting the training set of labeled pairs to make independent local models, as presented in Section \ref{sec:local}, prevents any sharing of information between different local models. Using a slightly different inference paradigm, one could argue that if a pair $(v,v')$ is known to be connected, and if both $v$ is similar to $v''$ and $v'$ is similar to $v'''$ in terms of features, then the pair $(v'',v''')$ is likely to be connected as well. Such induction principle is not possible with local models, since the pair $(v,v')$ is only considered by the local model of $v$, while $(v'',v''')$ is only considered by the local model of $v''$.

\begin{figure}[ht]
\begin{center}
\includegraphics[width=\textwidth]{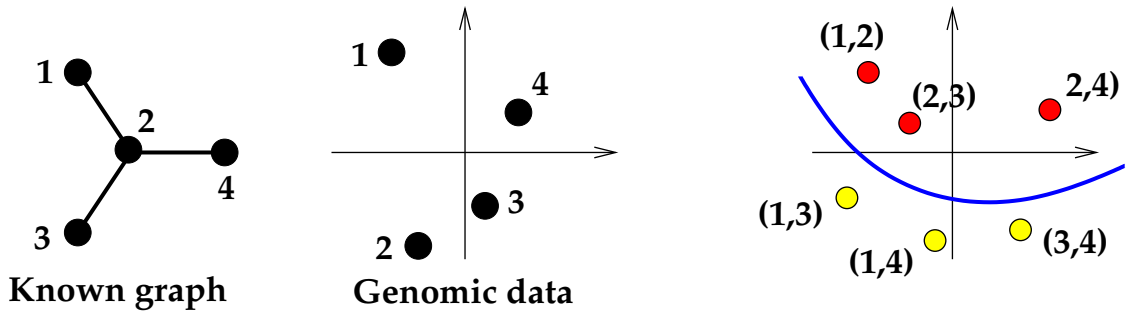}
\caption{With global models, we want to formulate the problem of edge prediction as a binary classification problem over pairs of vertices. A pair can be connected (label $+1$) or not connected (label $-1$). However the data available are attributes about each individual vertices (central picture). Hence we need to define a representation for pairs of vertices, as illustrated on the right-hand picture, in order to apply classical pattern recognition methods to discriminate between interacting and non-interacting pairs in the graph shown in the left-hand picture.}
\label{fig:globalproblem}
\end{center}
\end{figure}

In order to implement this inference paradigm, we need to work directly with pairs of vertices as patterns, and in particular to be able to represent any pair $(u,v)\in V\times V$ by a feature vector which we denote $\psi(u,v)$. As we originally have only data to characterize each individual protein $v$ by a vector $\phi(v)$, we therefore need to clarify how to derive a vector for a pair $\psi(u,v)$ from the vectors $\phi(u)$ and $\phi(v)$ that characterize $u$ and $v$. This problem is illustrated in Figure \ref{fig:globalproblem}.

As suggested in Section \ref{sec:pattrec}, kernels offer various useful tricks to design features, or equivalently kernels, for pairs of vertices starting from features for individual vertices. Let us consider for example a simple, although not very useful, trick to design a vector representation for a pair of vertices from a vector representation of individual vertices. If each vertex $v$ is characterized by a vector of features $\phi(v)$ of dimension $p$, we can choose to represent a pair of vertices $(u,v)$ by the concatenation of the vectors $\phi(u)$ and $\phi(v)$ into a single vector $\psi_\oplus(u,v)$ of size $2p$. In other words, we could consider their \emph{direct sum} defined as follows:
\begin{equation}\label{eq:directsum}
\psi_{\oplus}(u,v) = \phi(u)\oplus\phi(v) = \left(\begin{array}{c}\phi(u) \\\phi(v)\end{array}\right)\,.
\end{equation}
If the dimension $p$ is large, one can avoid the burden of computing and storing large-dimensional vectors by using the kernel trick. Indeed, let us denote by $K_{V}$ the kernel for vertices induced by the vector representation $\phi$, namely, $K_{V}(v,v') = \phi(v)^\top\phi(v')$ for any pair of vertices $(v,v')$, and let us assume that $K_{V}(v,v')$ can be easily computed. Then the following computation, similar to (\ref{eq:kernelsum}), shows that the kernel $K_\oplus$ between two pairs of vertices $(a,b)$ and $(c,d)$ induced by the vector representation $\psi_{\oplus}$ is easily computable as well: 
\begin{equation}
\begin{split}
K_{\oplus}\br{(a,b),(c,d)} &= \psi_{\oplus}(a,b)^\top\psi_{\oplus}(c,d) \\
& = \left(\begin{array}{c}\phi(a) \\\phi(b)\end{array}\right)^\top \left(\begin{array}{c}\phi(c) \\\phi(d)\end{array}\right) \\
&= \phi(a)^\top \phi(c) + \phi(b)^\top\phi(d) \\
& = K_{V}(a,c) + K_{V} (b,d)\,.
\end{split}
\end{equation}
Hence the kernel between pairs is here simply obtained by summing individual kernels, and an algorithm like an SVM could be trained on the original training set of labeled pairs, to predict the label of new pairs not in the training set. Although attractive at first sight, this formulation has an important limitation. Training an SVM (or any linear classifier) means that one estimates a linear function in the space of direct sums, i.e., a function for pairs of the form: $h(u,v) = w^\top\psi_\oplus(u,v)$. The vector $w$ (of size $2p$) can be decomposed as a concatenation of two parts $w_{1}$ and $w_{2}$ of size $p$, i.e., $w=w_{1}\oplus w_{2}$. We can then rewrite the linear function as:
$$
h(u,v) = \br{w_{1}\oplus w_{2}}^\top \br{\phi(u)\oplus\phi(v)} = w_{1}^\top \phi(u) + w_{2}\top \phi(v)\,.
$$
Hence any linear classifier $h(u,v)$ in the space defined by the direct sum representation decomposes as a sum of two independent functions:
$$
h(u,v) = h_{1}(u) + h_{2}(v)\,,
$$
with $h_{i}(v)=w_{i}^\top v$ for $i=1,2$. This is in general an unfortunate property since it implies, for example, that whatever the target vertex $u$, if we sort the candidate vertices $v$ that can interact with $u$ according to the classifier (i.e., if we rank $v$ according to the value of $h(u,v)$), then the order will not depend on $u$. In other words, each vertex $v$ would be associated to a particular score $h_{2}(v)$ that could be thought of as its general propensity to interact, and the prediction of vertices connected to a particular vertex $u$ would only depend on the scores of the vertices tested, not on $u$ itself. This clearly limits the scope of the classification rules that linear classifiers can produce with the direct sum representations, which suggests that this approach should not be used in general.

A generally better alternative to the direct sum $\psi_\oplus(u,v)$ is to represent a pair of vertices $(u,v)$ by their \emph{direct product}:
\begin{equation}\label{eq:tensorproduct}
\psi_{\otimes}(u,v) = \phi(u)\otimes\phi(v)\,.
\end{equation}
If $\phi(u)$ and $\phi(v)$ each has a dimension $p$, then the direct product $\psi_{\otimes}(u,v)$ is by definition a vector of dimension $p^2$ whose entries are all possible products between a feature of $\phi(u)$ and a feature of $\phi(v)$. An interesting property of the direct product is that it encodes features that are characteristic of the pair $(u,v)$, and not merely of $u$ and $v$ taken separately. For example, let us assume that $\phi(u)$ and $\phi(v)$ contain binary features that indicate the presence or absence of particular features in $u$ and $v$. Then, because the product of binary features is equivalent to a logical $AND$, the vector $\psi_{\otimes}(u,v)$ contains binary features that indicate the joint occurrence of particular pairs of features in $u$ and $v$. As a result, contrary to the direct sum representation $\psi_{\oplus}(u,v)$, linear classifiers in the space defined by $\psi_{\otimes}(u,v)$ could predict that $a$ is more likely to interact with $u$ than $b$, while $b$ is more likely to interact with $v$ than $a$, for two different target vertices $u$ and $v$.

The price to pay in order to obtain this large flexibility is that the dimension of the representation, namely $p^2$, can easily get very large. Typically, if an individual gene is characterized by a vector of dimension $1,000$ to encode expression data, phylogenetic profiles and/or subcellular localization information, then the direct product representation has one million dimensions. Such large dimensions may cause serious problems in terms of computation time and memory storage for practical applications. Fortunately, if one works with kernel methods like SVM, a classical trick allows to compute efficiently the inner product between two tensor product vectors from the inner products between individual vectors:
\begin{equation}\label{eq:productkernel}
\begin{split}
K_{\otimes}\br{(a,b),(c,d)} &= \psi_{\otimes}(a,b)^\top\psi_{\otimes}(c,d) \\
& = \br{\phi(a)\otimes\phi(b)}^\top\br{\phi(c)\otimes\phi(d)} \\
&= \phi(a)^\top \phi(c) \times \phi(b)^\top\phi(d) \\
& = K_{V}(a,c) \times K_{V} (b,d)\,,
\end{split}
\end{equation}
where the third line is a classical result easily demonstrated by expanding the inner product between tensor product vectors. Hence one obtains the kernel between two pairs of vertices by just multiplying together the kernel values involving each vertex of the first pair and the corresponding vertex of the second pair.

The direct sum (\ref{eq:directsum}) and product (\ref{eq:tensorproduct}) representations correspond to representations of ordered paired, which usually map a pair $(u,v)$ and its reverse $(v,u)$ to different vectors. For example, the concatenation of two vectors $\phi(u)$ and $\phi(v)$ is generally different from the concatenation of $\phi(v)$ and $\phi(u)$, i.e., $\psi_\oplus(u,v) \neq \psi_\oplus(v,u)$, except when $\phi(u)=\phi(v)$. Hence these representations are well adapted to the prediction of edges in directed graphs, where an ordered pair $(u,v)$ can represent an edge form $u$ to $v$ and the pair $(v,u)$ then represents the different edge from $v$ to $u$. When the graph of interest is not directed, then it can be advantageous to also represent an undirected pair $\cbr{u,v}$. An extension of the tensor product representation was for example proposed by \cite{Ben-Hur2005Kernel} with the following \emph{tensor product pairwise kernel (TPPK)} representation for undirected pairs:
\begin{equation}\label{eq:psitppk}
\psi_{TPPK}\br{\cbr{u,v}} = \psi_{\otimes}(u,v) + \psi_{\otimes}(v,u)\,.
\end{equation}
This representation is the symetrized version of the direct product representation, which makes it invariant to a permutation in the order of the two vertices in a pair. The corresponding kernel is easily derived as follows:
\begin{equation}\label{eq:tppk}
\begin{split}
K_{TPPK}\br{\cbr{a,b},\cbr{c,d}} &= \psi_{TPPK}(\cbr{a,b})^\top\psi_{TPPK}(\cbr{c,d}) \\
& = \br{\psi_{\otimes}(a,b)+\psi_{\otimes}(b,a)}^\top\br{\psi_{\otimes}(c,d) + \psi_{\otimes}(d,c)} \\
&= \psi_{\otimes}(a,b)^\top\psi_{\otimes}(c,d) + \psi_{\otimes}(a,b)^\top\psi_{\otimes}(d,c) \\
& \quad + \psi_{\otimes}(b,a)^\top\psi_{\otimes}(c,d) +\psi_{\otimes}(b,a)^\top\psi_{\otimes}(d,c) \\
&= 2 \cbr{ K_{V}(a,c)K_{V}(b,d) + K_{V}(a,d)K_{V}(b,c)}\,.
\end{split}
\end{equation}
Once again we see that the inner product in the space of the TPPK representation is easily computed from the values of kernels between individual vertices, without the need to compute explicitly the $p^2$-dimension TPPK vector. This approach is therefore, again, particularly well suited to be used in combination with an SVM or any other kernel method.

An alternative and perhaps more intuitive justification for the TPPK kernel (\ref{eq:tppk}) is in terms of similarity or distance between pairs induced by this formulation. Indeed, when a kernel $K_{V}$ is such that $K_{V}(v,v)=1$ for all $v$, which equivalently means that all vectors $\phi(v)$ are normalized to unit norm, then the value of the kernel $K_{V}(u,v)$ is a good indicator of the ``similarity'' between $u$ and $v$. In particular we easily show in that case that:
$$
K_{V}(u,v) = \phi(u)^\top\phi(v) = 1 - \frac{||\phi(u) - \phi(v)||^2}{2}\,,
$$
which shows that $K_{V}(u,v)$ is ``large'' when $\phi(u)$ and $\phi(v)$ are close to each other, i.e., when $u$ and $v$ are considered ``similar''. An interesting point of view to define a kernel over pairs in this context is then to express it in terms of similarity: when do we want to say that an unordered pair $\cbr{a,b}$ is similar to a pair $\cbr{c,d}$, given the similarities between individual vertices? One attractive formulation is to consider them similar if either (i) $a$ is similar to $c$ and $b$ is similar to $d$, or (ii) $a$ is similar to $d$ and $b$ is similar to $c$. Translating these notions into equation, the TPPK kernel formulation (\ref{eq:tppk}) can be thought of as an implementation of this principle \cite{Ben-Hur2005Kernel}.

At this point, it is worth mentioning that although the tensor product (\ref{eq:tensorproduct}) for directed pairs, and its extension (\ref{eq:psitppk}) for undirected pairs, can be considered as ``natural'' default choices to represent pairs of vertices as vectors from representations of individual vertices, they are by no means the only possible choices. As an example, let us briefly mention the construction of \cite{Vert2007new} who propose to represent an undirected pair as follows:
\begin{equation}\label{eq:psimlpk}
\psi_{MLPK}\br{u,v} = \br{\phi(u)-\phi(v)}^{\otimes 2} = \br{\phi(u)-\phi(v)}\otimes \br{\phi(u)-\phi(v)}\,.
\end{equation}
The name MLPK stands for \emph{metric learning pairwise kernel}. Indeed, \cite{Vert2007new} shows that training a linear classifier in the representation defined by the MLPK vector (\ref{eq:psimlpk}) is equivalent, in some situations, to estimating a new metric in the space of individual vertices $\phi(v)$, and classifying a pair as positive or negative depending on whether or not the distance between $\phi(u)$ and $\phi(v)$ (with respect to the new metric) is below a threshold or not. Hence this formulation can be particularly relevant in cases where connected vertices seem to be ``similar'', in which case a linear classifier coupled with the MLPK representation can learn by itself the optimal notion of ``similarity'' that should be used in a supervised framework. For example, if a series of expression values for genes across a range of experiments is available, one could argue that proteins coded by genes with ``similar'' expression profiles are more likely to interact than others, and therefore that a natural way to predict interaction would be to measure a ``distance'' between all pairs of expression profiles and threshold it above some value to predict interactions. The question of how to chose a ``distance'' between expression profiles is then central, and instead of choosing \emph{a priori} a distance such as the Euclidean norm, one could typically let an SVM train a classifier with the MLPK representation to mimic the process of choosing an optimal way to measure distances in order to predict interactions.

An interesting property of the MLPK representation (\ref{eq:psimlpk}) is that, as for the tensor product and TPPK representation, it leads to an inner product that can easily be computed without explicitly computing the $p^2$-dimensional vector $\phi_{MLPK}(a,b)$:
\begin{equation}\label{eq:mlpk}
\begin{split}
K_{MLPK}\br{\cbr{a,b},\cbr{c,d}}
&= \psi_{MLPK}\br{a,b}^\top\psi_{MLPK}\br{c,d}\\
&= \sqb{\br{\phi(a)-\phi(b)}^{\otimes 2}}^\top \sqb{\br{\phi(c)-\phi(d)}^{\otimes 2}} \\
&= \sqb{\br{\phi(a)-\phi(b)}^\top\br{\phi(c)-\phi(d)}}^2 \\
&=\sqb{\phi(a)^\top\phi(c) - \phi(a)^\top\phi(d) -\phi(b)^\top\phi(c) + \phi(b)^\top\phi(d)}^2\\
&= \sqb{K_{V}(a,c) - K_{V}(a,d) - K_{V}(b,c) + K_{V}(b,d)}^2\,.
\end{split}
\end{equation}

\subsection{Remarks}
We have shown how the general problem of graph reconstruction can be formulated as a pattern recognition problem (Sections \ref{sec:problem}-\ref{sec:graphpatrec}), and described several instances of this idea: either by training a multitude of local models to learn the local structure of the graph around each node (Section \ref{sec:local}), which boils down to a series of pattern recognition problems over vertices, or by training a single global model to predict whether any given pair of vertices interacts or not, which requires the definition of a vector representation (or equivalently of a kernel) for pairs of vertices (Section \ref{sec:global}). Our presentation has been fairly general, in order to highlight the general ideas behind the approach and the main choices one has to make in order to implement it. Now, we discuss several important questions that one must also address to implement the idea on any particular problem.
\begin{itemize}
\item \emph{Directed or undirected graph}. As pointed out in the introduction, some biological networks are better represented by undirected graphs (e.g., the PPI network) while other are more naturally viewed as directed graphs (e.g., a gene regulatory network). In the course of our presentation we have shown that some methods are specifically adapted to one case or the other. For example, the MLPK and TPPK kernel formulations to learn global models (equations \ref{eq:tppk} and \ref{eq:mlpk}) are specifically tailored to solve problems over undirected pairs, i.e., to reconstruct undirected graphs. On the other hand, the local models (Section \ref{sec:local}) or the global models with the direct product kernel (\ref{eq:productkernel}) are naturally suited to infer interactions between directed pairs, i.e., to reconstruct directed graphs. However, one can also use them to reconstruct undirected graph by simply counting each undirected pair $\cbr{u,v}$ as two directed pairs $(u,v)$ and $(v,u)$. In the training step, this means that we can replace each labeled undirected pair (i.e., undirected edge known to be present or absent) by two directed pairs labeled by the same label. In the prediction step, this means that one would get a prediction for the pair $(u,v)$ and another prediction for the pair $(v,u)$, that have no reason to be consistent between each other to predict whether the undirected pair $\cbr{u,v}$ is connected or not. In order to reconcile both predictions, one typically can take the average of the prediction scores of the classifiers for both directed pairs in order to make a unique prediction score for the undirected pair.
\item \emph{Different types of edges}. Some biological networks are better represented by graphs with edges having additional attributes, such as a label among a finite set of possible labels. For example, to describe a gene regulatory network it is common to consider two types or regulations (edges), namely activation or inhibition. In terms of prediction, this means that we not only need to predict whether two vertices are connected or not, but also by what type of edges they are connected. A simple strategy to extend the pattern recognition paradigm to this context is to see the problem not as a binary classification problem, but more generally as a multi-class classification problem. In the previous example, one should for example assign each pair $(u,v)$ to one of the three classes (no regulation, activation, inhibition). Luckily the extension of pattern recognition algorithms to the multi-class setting is a well-studied field in machine learning for which many solutions exist \cite{Hastie2001elements,Bishop2006Pattern}. For example, a popular approach to solve a classification problem with $k$ classes is to replace it by $k$ binary classification problems, where each binary problem discriminates versus data in one of the $k$ classes and the rest of the data. Once the $k$ classifiers are trained, they can be applied to score each new candidate point, and the class corresponding to the classifier that outputs the largest score is predicted. Other approaches also exist besides this scheme, known as the \emph{one-versus-all} strategy. Overall they show that the pattern recognition formulation can easily accommodate the prediction of different edge types just by using a multi-class classification algorithm.
\item \emph{Negative training pairs}. While most databases store information about the presence of edges and can be used to generate positive training examples, few if any negative interactions are usually reported. This is an important problem since, as we formulated it in Section \ref{sec:pattrec}, the typical pattern recognition formalism requires positive as well as negative training examples. In order to overcome this obstacle several strategies can be pursued. A first idea would be to refrain from focusing exclusively on pattern recognition algorithms which are not adapted to the lack of negative examples, and use instead algorithms specifically designed to handle only positive examples. For example, many methods in statistics for density estimation or outlier detection are designed to estimate a small region that contains all or most of the positive training points. If such a region of ``positive examples'' is found around pairs known to be connected, then a new pair of vertices can be predicted to be connected if it also lies in the region. An algorithm like the one-class SVM \cite{Scholkopf2001Estimating} is typically adapted to this setting, and can accommodate all the kernel formulations we presented so far. A second idea would be to keep using algorithms for binary classification, and \emph{generate} negative examples. Perhaps the simplest way to do this is to randomly sample pairs of vertices, among the ones not known to be connected, and declare that they are negative examples. As the graph is usually supposed to be sparse, most pairs of vertices randomly picked by this process indeed do not interact, and are correctly labeled as negative. On the other hand, the few pairs that would be wrongly labeled as negative with this procedure, namely the pairs that interact although we do not know it yet, are precisely the one we are interested to find. There may then be a danger that by labeling them as negative and training a classifier based on this label, we could have more difficulties to find them. To overcome this particular issue of generating false negative examples in the training set, one may again consider two ideas. First, try to reduce the quantity of wrongly labeled negative training pairs by, e.g., using additional sources of informations to increase the likelihood that they to not interact. For example, if one wants to choose pairs of proteins that are very unlikely to interact, he may restrict himself to proteins known to be located in different subcellular localization, which in theory prevent any possibility of physical interaction. While this may increase the size of the training set, there is also a danger to bias the training set towards "easy" negative examples \cite{Ben-Hur2006Choosing}. The second idea is to accept the risk of generating false negative training examples, but then to be careful at least that the predictive models never predict the label of a pair that was used during its training. This can be achieved, for example, by splitting the set of candidate negative pairs (i.e., those not known to interact) into $k$ disjunct subsets, train a classifier using $k-1$ of these subsets as negative training examples and using the resulting classifier to predict the labels of pairs in the subset that was left apart. Repeating this procedure $k$ times leads to the possibility of predicting the labels for the $k$ subsets, without ever predicting the label of a negative example that was used during training. This strategy was for example used in \cite{Mordelet2008SIRENE}.
\item \emph{Presence or absence of errors in the training data}. Besides the lack of known negative examples, one may also be confronted with possible errors in the positive training examples, i.e., false positives in the training set. Indeed, many databases of biological networks contain both certain interactions, and interactions believed to be true based on various empirical evidences but that could be wrong. This is particularly true, for example, for PPI networks when physical interactions have been observed with high-throughput technologies such as the yeast two-hybrid system, which is known to be prone to many false positive detections. In that case, we should not only be careful when using the data as positive training examples, but we may even consider the possibility of using the predictive algorithms to remove wrong positive annotations from the training set. Regarding the problem of training models with false positive training examples, this may not be a major obstacle since one of the strengths of statistical pattern recognition methods is precisely to accept ``noise'' or errors in the data. On the other hand, if one wants to further use the models to correct the training data, then a specific procedure could be imagined, for example similar to the procedure described in the previous paragraph to predict the label of false negative examples.
\end{itemize}

\section{Examples}\label{sec:exp}
Recently, the different approaches, surveyed in Section \ref{sec:pat}, have been extensively tested and compared to other approaches in several publications. In this section, we review the main findings of these publications, focusing on our three running examples of biological networks.

\subsection{Reconstruction of a metabolic network}\label{sec:metabolic}

The reconstruction of metabolic networks has been among the first applications that motivated the line of research surveyed in this chapter \cite{Yamanishi2004Protein,Vert2005Supervised,Yamanishi2005Supervised, Bleakley2007Supervised}. We consider here the problem of inferring the metabolic gene network of the yeast {\it S. cerevisiae} with the enzymes represented as vertices, and an edge between two enzymes when the
two enzymes catalyse successive reactions. The dataset, proposed by \cite{Yamanishi2005Supervised}, consists of $668$ vertices
(enzymes) and 2782 edges between them which were extracted from the 
KEGG database of metabolic pathways \cite{Kanehisa2004KEGG}. In order to predict edges in these networks, \cite{Bleakley2007Supervised} used various genomic datasets and compared different inference methods. Following \cite{Yamanishi2005Supervised}, the data used to characterize enzymes comprise 157 expression data measured under different experimental conditions \cite{Eisen1998Cluster,Spellman1998Comprehensive},  a vector of $23$ bits representing
the localization of the enzymes (found or not found) in $23$ locations in the cell
 determined experimentally \cite{Huh2003Global}, and the phylogenetic profiles 
of the enzymes  as vectors of $145$ bits denoting the presence or
absence of the enzyme in $145$ fully sequenced genomes \cite{Kanehisa2004KEGG}. Each type of data was processed and transformed into a kernel as described in \cite{Yamanishi2005Supervised,Kato2005Selective}, and all matrices were summed together to produce a single kernel integrating heterogeneous data.

On a common $5$-fold cross-validation setting, \cite{Bleakley2007Supervised} compared different methods including local models (Section \ref{sec:local}), the TPPK and MLPK kernels (Section \ref{sec:global}) as well as several other methods: a direct \emph{ de novo} approach which only infers edges between similar vertices, an approach based on kernel canonical correlation analysis (KCCA) \cite{Yamanishi2004Protein}, and a matrix completion algorithm based on an \emph{em} procedure \cite{Tsuda2003em,Kato2005Selective}. On each fold of the cross-validation procedure, each method uses the training set to learn a model and makes predictions on pairs in the test set. All methods associate a score to all pairs in the test set, hence by thresholding this score at different levels they can predict more or less edges. Results were assessed in terms of average ROC curve (which plots the percentage of true positives as a function of the percentage of false positives, when the threshold level is varied) and average precision/recall curve (which plots the percentage of true positives among positive predictions, as a function of the percentage of true positives among all positives). In practical applications, the later criteria is a better indicator of the relevance of a method than the former one. Indeed, as biological networks are usually sparse, the number of negatives far exceeds the number of positives, and only large precision (over a recall as large as possible) can be tolerated if further experimental validations are expected.

\begin{figure}[ht]
\begin{center}
\includegraphics[width=0.45\textwidth]{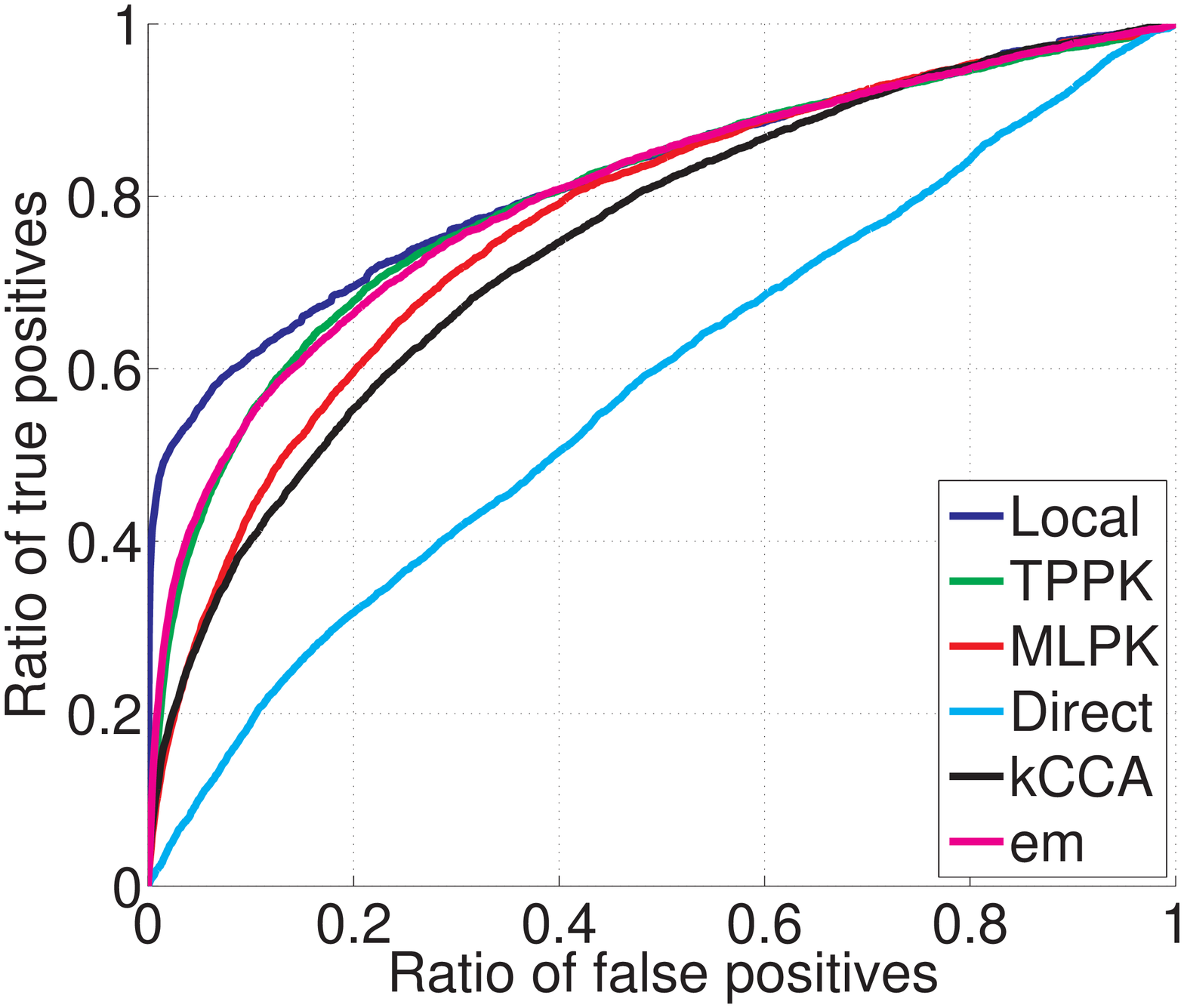}
\includegraphics[width=0.45\textwidth]{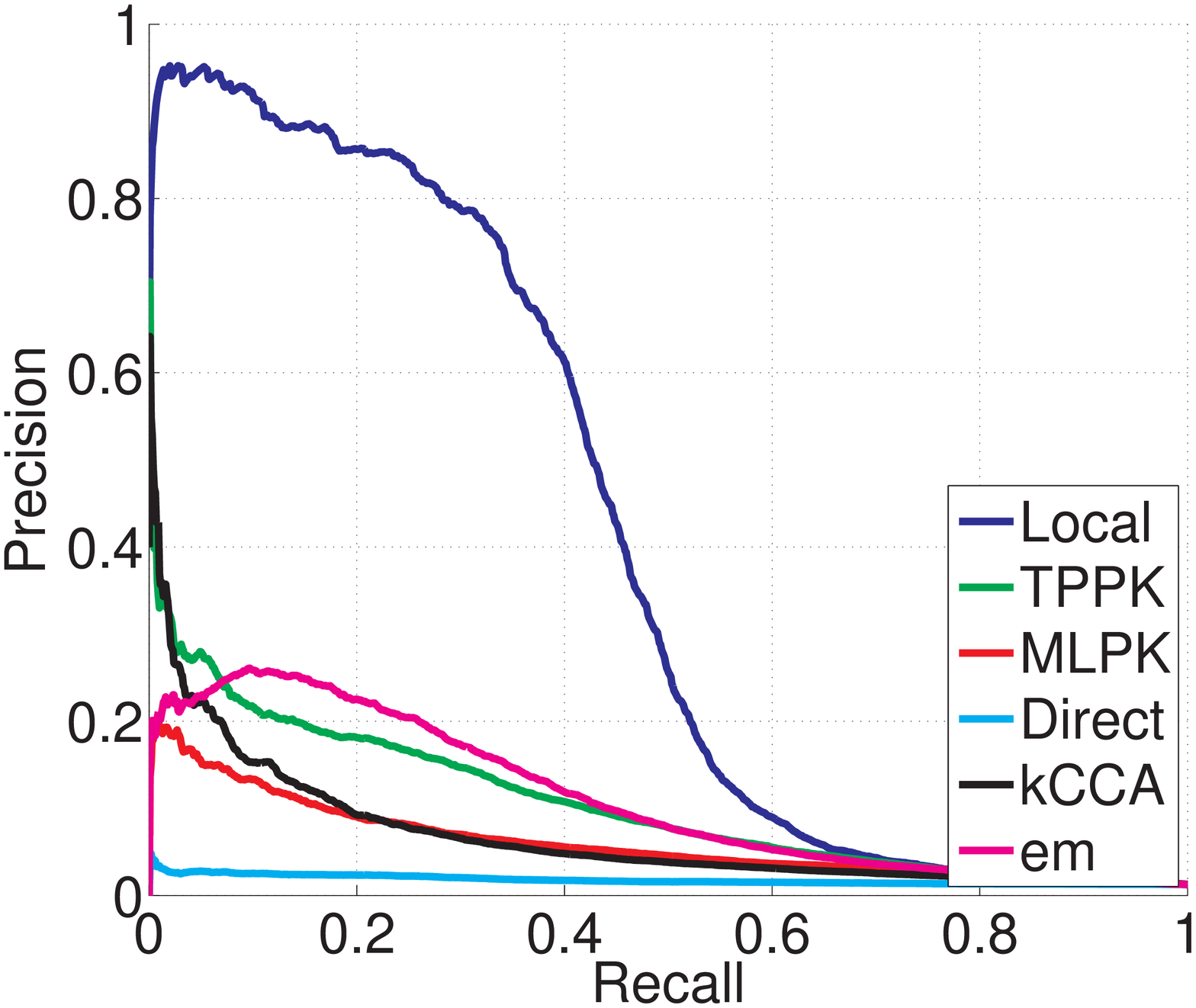}
\caption{Performance of different methods for the reconstruction of metabolic networks (from \cite{Bleakley2007Supervised}): ROC (left) and precision/recall (right) curves.}
\label{fig:metabolic}
\end{center}
\end{figure}

Figure \ref{fig:metabolic} shows the performance of the different methods on this benchmark. A very clear advantage for the local model can be seen. In particular it is the only method tested that can produce predictions at more than 80$\%$ precision. There is no clear winner among the other supervised methods, while the direct approach which is the only \emph{de novo} method in this comparison, is clearly below the supervised methods.

\subsection{Reconstruction of a PPI network}
As a second application, we consider the problem of inferring missing edges in the PPI network of the yeast \emph{S. cerevisiae}.  The gold standard PPI graph used to perform a cross-validation experiment is a set of high-confidence interactions supported by several experiments provided by  \cite{Mering2002Comparative} and also used in \cite{Kato2005Selective}. After removal of proteins without interactions we end up with a graph involving 2438 interactions (edges) among 984 proteins (vertices). In order to reconstruct missing edges the genomic data used are the same as those used for the reconstruction of the metabolic network in Section \ref{sec:metabolic}, namely gene expression, protein localization and phylogenetic profiles, together with a set of yeast two-hybrid data obtained from \cite{Ito2001comprehensive} and \cite{Uetz2000comprehensive}. The later was converted into a positive definite kernel using a diffusion kernel, as explained in  \cite{Kato2005Selective}. Again, all datasets were combined into a unique kernel by adding together the four individual kernels.

\begin{figure}[ht]
\begin{center}
\includegraphics[width=0.45\textwidth]{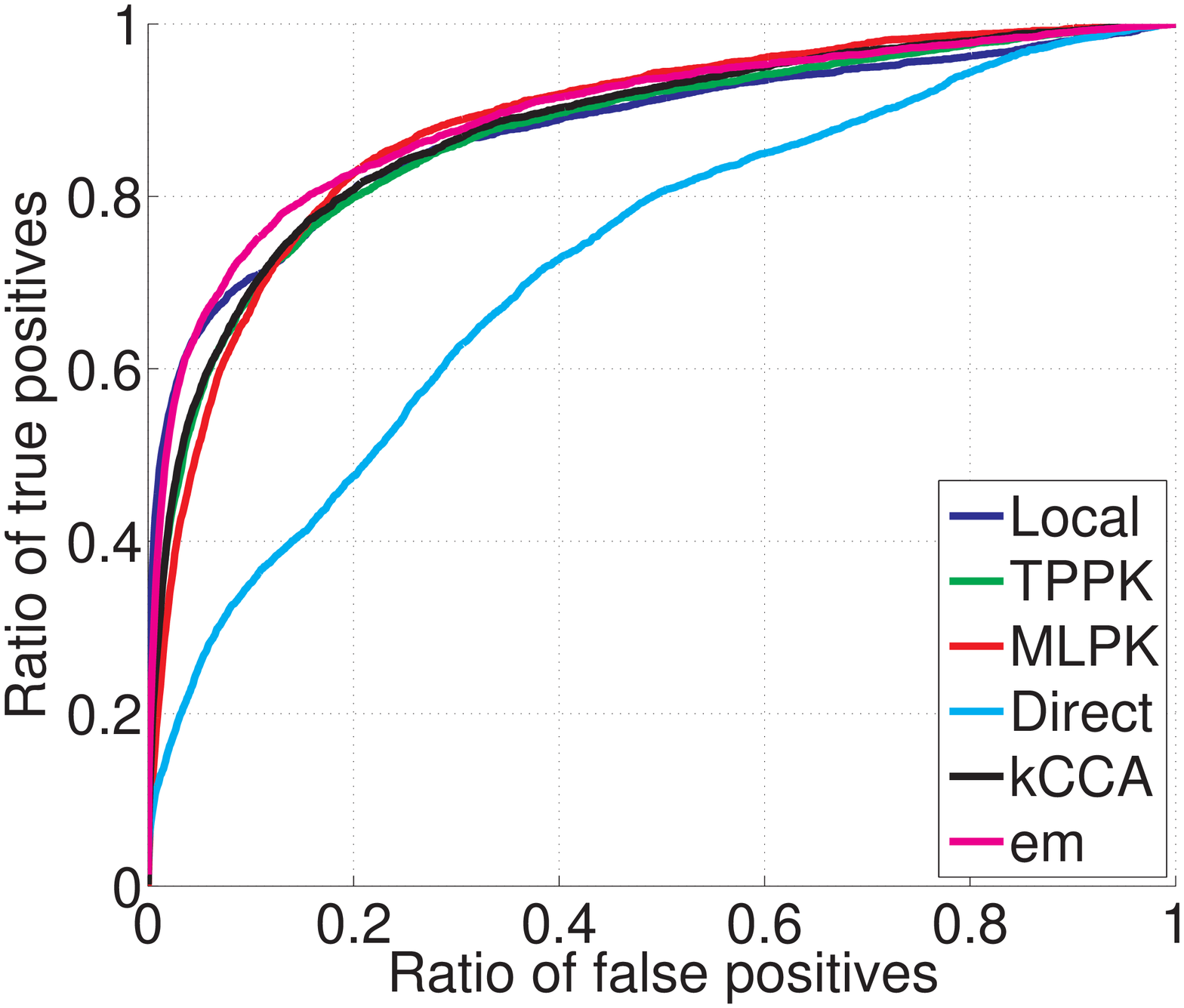}
\includegraphics[width=0.45\textwidth]{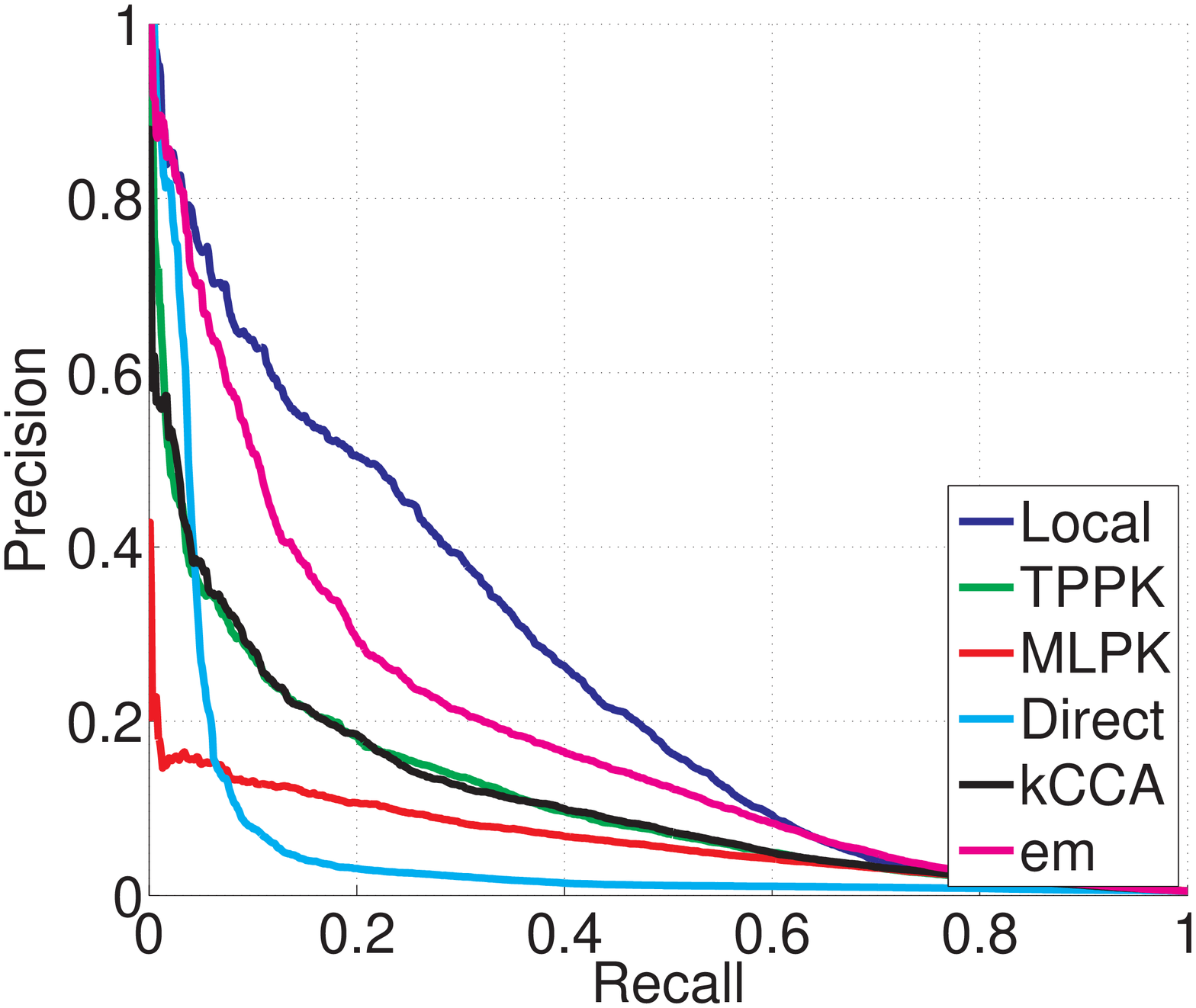}
\caption{Performance of different methods for the reconstruction of the PPI network (from \cite{Bleakley2007Supervised}): ROC (left) and precision/recall (right) curves.}
\label{fig:ppi}
\end{center}
\end{figure}

Figure \ref{fig:ppi} shows the performances of the different methods, using the same experimental protocol as the one used for the experiment with metabolic network reconstruction in Section \ref{sec:metabolic}. Again the best method is the local model, although it outperforms the other methods with a smaller margin than for the reconstruction of the metabolic network (Figure \ref{fig:metabolic}). Again the ROC curve of the \emph{de novo} direct method is clearly below the curves of the supervised methods, although this time it leads to large precision at low recall. This means that a few interacting pairs can very easily be detected because they have very similar genomic data.

\subsection{Reconstruction of gene regulatory networks}
Finally, we report the results of an experiment conducted for the inference of a gene regulatory network by \cite{Mordelet2008SIRENE}. In that case the edges between transcription factors and the genes they regulate are directed, therefore only the local model of Section \ref{sec:local} is tested. It is compared to a panel of other state-of-the-art methods dedicated to the inference of gene regulatory networks from a compendium of gene expression data, using a benchmark proposed by \cite{Faith2007Large-scale}. More precisely, the goal of this experiment is to predict the regulatory network of the bacteria \emph{E. coli} from a compendium of 445 microarray expression profiles for 4345 genes. The microarray were collected under different experimental conditions such as PH changes, growth phases, antibiotics, heat shock, different media, varying oxygen concentrations and numerous genetic perturbations. The goal standard graph used to assess the performance of different methods by cross-validation consists of 3293 experimentally confirmed regulations between 154 TF and 1211 genes, extracted from the RegulonDB database \cite{Salgado2006RegulonDB}.

\begin{figure}[ht]
\begin{center}
\includegraphics[width=0.45\textwidth]{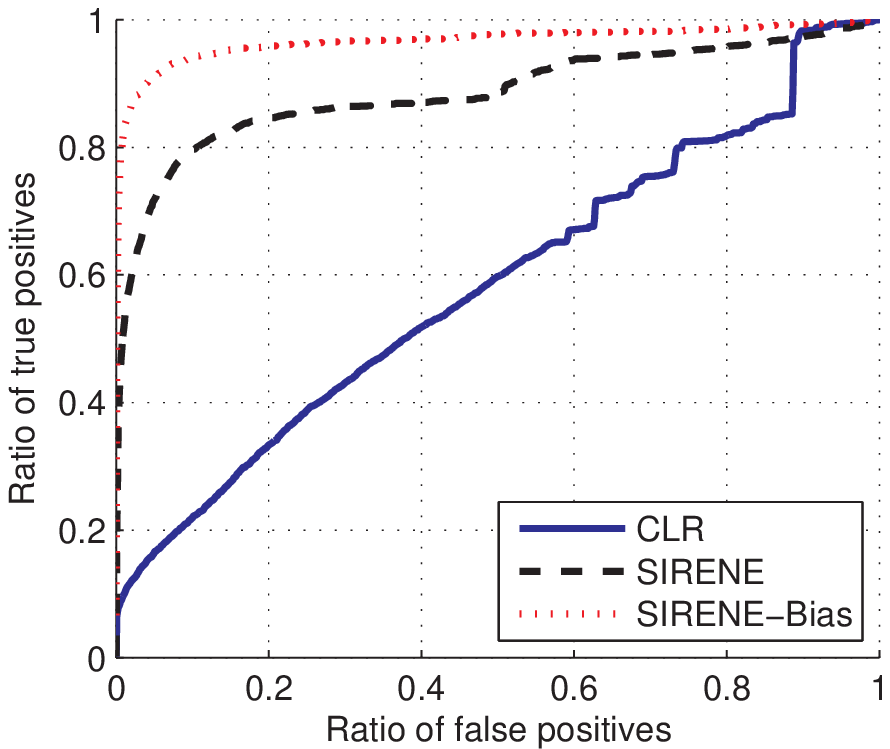}
\includegraphics[width=0.45\textwidth]{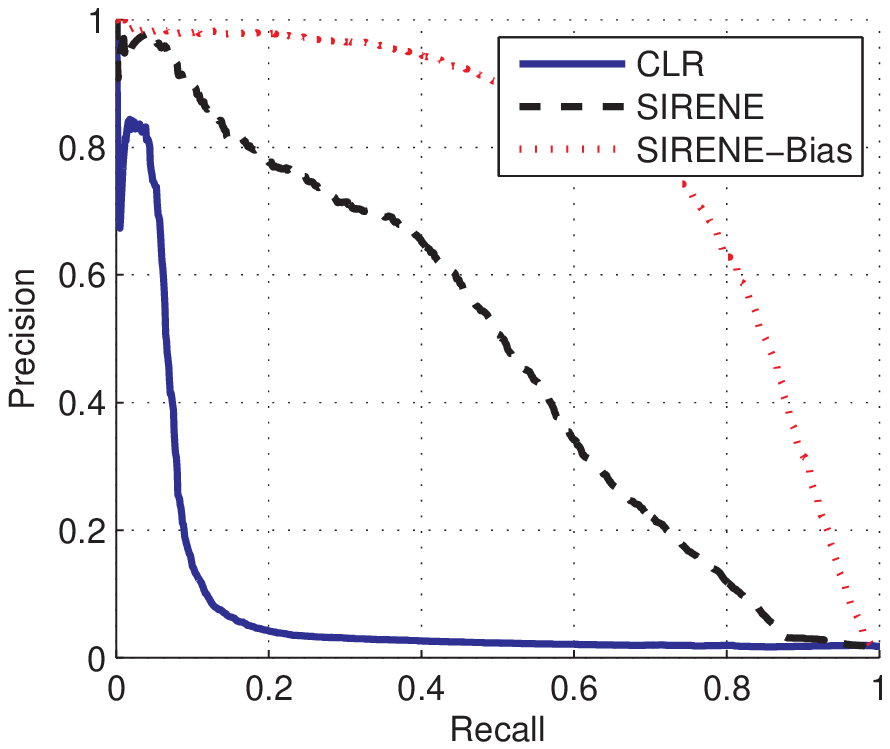}
\caption{Comparison of the CLR method and the local pattern recognition approach (called \emph{SIRENE)} on the reconstruction of a regulatory network: ROC (left) and precision/recall (right) curves. The curve \emph{SIRENE-Bias} corresponds to the performance of \emph{SIRENE} with a cross-validation procedure which does not take into account the organization of genes in operons, thus introducing an artificial positive bias in the result.}
\label{fig:sirene}
\end{center}
\end{figure}

In \cite{Faith2007Large-scale} this benchmark was used to compare different algorithms, including Bayesian networks \cite{Friedman2000Using}, ARACNe \cite{Margolin2006ARACNE}, and the context likelihood of relatedness (CLR) algorithm \cite{Faith2007Large-scale}, a new method that extends the relevance networks class of algorithms \cite{Butte2000Discovering}. They observed that CLR outperformed all other methods in prediction accuracy, and experimentally validated some predictions. CLR can therefore be considered as state-of-the-art among methods that use compendia of gene expression data for large-scale inference of regulatory networks. However, all the methods compared in \cite{Faith2007Large-scale} are \emph{de novo}, and the goal of \cite{Mordelet2008SIRENE} was to compare the supervised local approach to the best \emph{de novo} method on this benchmark, namely the CLR algorithm. Using a $3$-fold cross-validation procedure (see details in \cite{Mordelet2008SIRENE}), they obtained the curves in Figure \ref{fig:sirene}.
We can observe that the local supervised approach (called \emph{SIRENE} for \emph{Supervised Inference of REgulatory NEtwork}) strongly outperforms the CLR method on this benchmark.  The recall obtained by SIRENE, i.e., the proportion of known regulations that are correctly predicted, is several times larger than the recall of CLR at all levels of precision. More precisely, Table \ref{tab:results} compares the recalls of SIRENE, CLR and several other methods at $80\%$ and $60\%$ precision. The other methods reported are relevance network \cite{Butte2000Discovering}, ARACNe \cite{Margolin2006ARACNE}, and a Bayesian network \cite{Friedman2000Using} implemented by \cite{Faith2007Large-scale}.
\begin{table}[htdp!]
\caption{Recall of different gene regulation prediction algorithm at different levels of precision ($60\%$ and $80\%$ (from \cite{Mordelet2008SIRENE}).}
\begin{center}
\begin{tabular}{|l|c|c|c|c|}
\hline
Method & Recall at 60\% & Recall at 80\%\\
\hline
SIRENE & \bf 44.5\% & \bf 17.6\% \\
\hline
CLR & 7.5\% & 5.5\% \\
\hline
Relevance networks & 4.7\% & 3.3\%  \\
\hline
ARACNe & 1\% & 0\%  \\
\hline
Bayesian network & 1\% & 0\%  \\
\hline
\end{tabular}
\end{center}
\label{tab:results}
\end{table}

This experiment also highlights the special care that must be taken when performing a cross-validation procedure, in particular to make sure that no artificial bias is introduced. The curve called \emph{SIRENE-bias} in Figure \ref{fig:sirene} corresponds to a normal $k$-fold cross-validation procedure, where the set of genes is randomly split into $k$ folds and each fold is used in turn as test set. In the case of regulation in bacteria like \emph{E. coli}, however, it is known that TFs can regulate groups of genes clustered together on the genome, called operons. Genes in the same operons are transcribed in the same messenger RNA, and have therefore very similar expression values across different experiments. If two genes within the same operon are split in a training and test set during cross-validation, then it will be very easy to recognize that the one in the test set has the same label as the one in the training set, which will artificially increase the accuracy of the method. Hence in this case it is important to make sure that, during the random split into $k$ subsets, all genes within an operon belong to the same fold. The curve names \emph{SIRENE} in Figure \ref{fig:sirene} has been obtained with this unbiased procedure. The important difference between both curves highlights the importance of the bias induced by splitting operons in the cross-validation procedure.

\section{Discussion}\label{sec:discussion}
We reviewed several strategies to cast the problem of graph inference as a classical supervised classification problem, which can be solved by virtually any pattern recognition algorithm. Contrary to \emph{de novo} approaches, these strategies assume that a set of edges is already known and use the data available about vertices and known edges to infer missing edges. On several experiments involving the inference of metabolic, PPI and regulatory networks from a variety of genomic data, these methods were shown to give good results compared to state-of-the-art \emph{de novo} methods, and a particular implementation of this strategy (the local model) consistently gave very good results on all datasets.

In a sense the superiority of supervised methods over \emph{de novo} methods observed in the experiments is not surprising, because supervised methods use more informations. As in many real-world applications this additional information is available, it suggests that supervised methods may be a better choice than \emph{de novo} ones in many cases. It should be pointed out, though, that some of the methods we classified as \emph{de novo}, like for example Bayesian networks, could easily be adapted to the supervised inference scenario by putting constraints or prior distribution on the graph to be inferred. On the other hand, the strength of supervised methods depends critically on the availability of a good training set, which may not be available in some situations, such as inferring the structure of smaller graphs.

We observe that there is not a single way to cast the problem as a binary classification problem, which suggests that further research is needed to design optimally adapted methods. In particular, the local method, which performs best in the 3 benchmark experiments, has obvious limitations, such as its inability to infer new edges for vertices with no edge already known. The development of new strategies that keep the performance of the local methods for vertices with enough known edges, but borrow some ideas from, e.g., the global models of Section \ref{sec:global} to be able to infer edges for vertices with few or no known edge, is thus a promising research direction.


\begin{thebibliography}{10}

\bibitem{Akutsu2000Algorithms}
T.~Akutsu, S.~Miyano, and S.~Kuhara.
\newblock {A}lgorithms for identifying {B}oolean networks and related
  biological networks based on matrix multiplication and fingerprint function.
\newblock {\em J. Comput. Biol.}, 7(3-4):331--343, 2000.

\bibitem{Bansal2006Inference}
M.~Bansal, G.~Della~Gatta, and D.~Bernardo.
\newblock Inference of gene regulatory networks and compound mode of action
  from time course gene expression profiles.
\newblock {\em Bioinformatics}, 22(7):815--822, Apr 2006.

\bibitem{Ben-Hur2005Kernel}
A.~Ben-Hur and W.~S. Noble.
\newblock Kernel methods for predicting protein-protein interactions.
\newblock {\em Bioinformatics}, 21(Suppl. 1):i38--i46, Jun 2005.

\bibitem{Ben-Hur2006Choosing}
A.~Ben-Hur and W.~S. Noble.
\newblock Choosing negative examples for the prediction of protein-protein
  interactions.
\newblock {\em BMC Bioinformatics}, 7 Suppl 1:S2, 2006.

\bibitem{Bernardo2005Chemogenomic}
D.~Bernardo, M.~J. Thompson, T.~S. Gardner, S.~E. Chobot, E.~L. Eastwood, A.~P.
  Wojtovich, S.~J. Elliott, S.~E. Schaus, and J.~J. Collins.
\newblock Chemogenomic profiling on a genome-wide scale using
  reverse-engineered gene networks.
\newblock {\em Nat. Biotechnol.}, 23(3):377--383, Mar 2005.

\bibitem{Biau2006Statistical}
G.~Biau and K.~Bleakley.
\newblock Statistical inference on graphs.
\newblock {\em Statistics and Decisions}, 24(2):209--232, 2006.

\bibitem{Bishop2006Pattern}
C.M. Bishop.
\newblock {\em Pattern recognition and machine learning}.
\newblock Springer, 2006.

\bibitem{Bleakley2007Supervised}
K.~Bleakley, G.~Biau, and J.-P. Vert.
\newblock Supervised reconstruction of biological networks with local models.
\newblock {\em Bioinformatics}, 23(13):i57--i65, Jul 2007.

\bibitem{Butte2000Discovering}
A.~J. Butte, P.~Tamayo, D.~Slonim, T.~R. Golub, and I.~S. Kohane.
\newblock {D}iscovering functional relationships between {RNA} expression and
  chemotherapeutic susceptibility using relevance networks.
\newblock {\em Proc. Natl. Acad. Sci. USA}, 97(22):12182--12186, Oct 2000.

\bibitem{Chen2005stochastic}
K.-C. Chen, T.-Y. Wang, H.-H. Tseng, C.-Y.~F. Huang, and C.-Y. Kao.
\newblock {A} stochastic differential equation model for quantifying
  transcriptional regulatory network in {S}accharomyces cerevisiae.
\newblock {\em Bioinformatics}, 21(12):2883--2890, Jun 2005.

\bibitem{Chen1999Modeling}
T.~Chen, H.~L. He, and G.~M. Church.
\newblock {M}odeling gene expression with differential equations.
\newblock {\em Pac. Symp. Biocomput.}, pages 29--40, 1999.

\bibitem{Cristianini2000introduction}
N.~Cristianini and J.~Shawe-Taylor.
\newblock {\em An introduction to {S}upport {V}ector {M}achines and other
  kernel-based learning methods}.
\newblock Cambridge University Press, 2000.

\bibitem{Eisen1998Cluster}
M.~B. Eisen, P.~T. Spellman, P.~O. Brown, and D.~Botstein.
\newblock Cluster analysis and display of genome-wide expression patterns.
\newblock {\em Proc. {N}atl. {A}cad. {S}ci. {USA}}, 95:14863--14868, Dec 1998.

\bibitem{Faith2007Large-scale}
J.~J. Faith, B.~Hayete, J.~T. Thaden, I.~Mogno, J.~Wierzbowski, G.~Cottarel,
  S.~Kasif, J.~J. Collins, and T.~S. Gardner.
\newblock {L}arge-scale mapping and validation of {E}scherichia coli
  transcriptional regulation from a compendium of expression profiles.
\newblock {\em PLoS Biol.}, 5(1):e8, Jan 2007.

\bibitem{Friedman2000Using}
N.~Friedman, M.~Linial, I.~Nachman, and D.~Pe'er.
\newblock Using {B}ayesian {N}etworks to {A}nalyze {E}xpression {D}ata.
\newblock {\em J. {C}omput. {B}iol.}, 7(3-4):601--620, 2000.

\bibitem{Gardner2003Inferring}
T.~S. Gardner, D.~Bernardo, D.~Lorenz, and J.~J. Collins.
\newblock Inferring genetic networks and identifying compound mode of action
  via expression profiling.
\newblock {\em Science}, 301(5629):102--105, Jul 2003.

\bibitem{Hastie2001elements}
T.~Hastie, R.~Tibshirani, and J.~Friedman.
\newblock {\em The elements of statistical learning: data mining, inference,
  and prediction}.
\newblock Springer, 2001.

\bibitem{Huh2003Global}
W.-K. Huh, J.~V. Falvo, L.~C. Gerke, A.~S. Carroll, R.~W. Howson, J.~S.
  Weissman, and E.~K. O'Shea.
\newblock {G}lobal analysis of protein localization in budding yeast.
\newblock {\em Nature}, 425(6959):686--691, Oct 2003.

\bibitem{Ito2001comprehensive}
T.~Ito, T.~Chiba, R.~Ozawa, M.~Yoshida, M.~Hattori, and Y.~Sakaki.
\newblock A comprehensive two-hybrid analysis to explore the yeast protein
  interactome.
\newblock {\em Proc. {N}atl. {A}cad. {S}ci. {USA}}, 98(8):4569--4574, 2001.

\bibitem{Jansen2003Bayesian}
R.~Jansen, H.~Yu, D.~Greenbaum, Y.~Kluger, N.J. Krogan, S.~Chung, A.~Emili,
  M.~Snyder, J.F. Greenblatt, and M.~Gerstein.
\newblock A {B}ayesian networks approach for predicting protein-protein
  interactions from genomic data.
\newblock {\em Science}, 302(5644):449--453, 2003.

\bibitem{Jeong2001Lethality}
H.~Jeong, S.~P. Mason, A.-L. Barab{\'a}si, and Z.~N. Oltvai.
\newblock Lethality and centrality in protein networks.
\newblock {\em Nature}, 411:41--42, 2001.

\bibitem{Kanehisa2004KEGG}
M.~Kanehisa, S.~Goto, S.~Kawashima, Y.~Okuno, and M.~Hattori.
\newblock The {KEGG} resource for deciphering the genome.
\newblock {\em Nucleic {A}cids {R}es.}, 32(Database issue):D277--80, Jan 2004.

\bibitem{Kato2005Selective}
T.~Kato, K.~Tsuda, and K.~Asai.
\newblock {S}elective integration of multiple biological data for supervised
  network inference.
\newblock {\em Bioinformatics}, 21(10):2488--2495, May 2005.

\bibitem{Lanckriet2004statistical}
G.~R.~G. Lanckriet, T.~De~Bie, N.~Cristianini, M.~I. Jordan, and W.~S. Noble.
\newblock A statistical framework for genomic data fusion.
\newblock {\em Bioinformatics}, 20(16):2626--2635, 2004.

\bibitem{Marcotte1999Detecting}
E.M. Marcotte, M.~Pellegrini, H.-L. Ng, D.W. Rice, T.O. Yeates, and
  D.~Eisenberg.
\newblock Detecting {P}rotein {F}unction and {P}rotein-{P}rotein {I}nteractions
  from {G}enome {S}equences.
\newblock {\em Science}, 285:751--753, 1999.

\bibitem{Margolin2006ARACNE}
A.~A. Margolin, I.~Nemenman, K.~Basso, C.~Wiggins, G.~Stolovitzky,
  R.~Dalla~Favera, and A.~Califano.
\newblock {ARACNE}: an algorithm for the reconstruction of gene regulatory
  networks in a mammalian cellular context.
\newblock {\em BMC Bioinformatics}, 7 Suppl 1:S7, 2006.

\bibitem{Mordelet2008SIRENE}
F.~Mordelet and J.-P. Vert.
\newblock Sirene: Supervised inference of regulatory networks.
\newblock {\em Bioinformatics}, 2008.
\newblock In press.

\bibitem{Okamoto2007Prediction}
S.~Okamoto, Y.~Yamanishi, S.~Ehira, S.~Kawashima, K.~Tonomura, and M.~Kanehisa.
\newblock Prediction of nitrogen metabolism-related genes in anabaena by
  kernel-based network analysis.
\newblock {\em Proteomics}, 7(6):900--909, Mar 2007.

\bibitem{Pavlidis2002Learning}
P.~Pavlidis, J.~Weston, J.~Cai, and W.S. Noble.
\newblock Learning {G}ene {F}unctional {C}lassifications from {M}ultiple {D}ata
  {T}ypes.
\newblock {\em J. {C}omput. {B}iol.}, 9(2):401--411, 2002.

\bibitem{Pazos2001Similarity}
F.~Pazos and A.~Valencia.
\newblock Similarity of phylogenetic trees as indicator of protein-protein
  interaction.
\newblock {\em Protein {E}ng.}, 9(14):609--614, 2001.

\bibitem{Salgado2006RegulonDB}
H.~Salgado, S.~Gama-Castro, M.~Peralta-Gil, E.~D{\'i}az-Peredo,
  F.~S{\'a}nchez-Solano, A.~Santos-Zavaleta, I.~Mart{\'i}nez-Flores,
  V.~Jim{\'e}nez-Jacinto, C.~Bonavides-Mart{\'i}nez, J.~Segura-Salazar,
  A.~Mart{\'i}nez-Antonio, and J.~Collado-Vides.
\newblock {RegulonDB} (version 5.0): {Escherichia coli K-12} transcriptional
  regulatory network, operon organization, and growth conditions.
\newblock {\em Nucleic Acids Res.}, 34(Database issue):D394--D397, Jan 2006.

\bibitem{Scholkopf2001Estimating}
B.~Sch{\"o}lkopf, J.~C. Platt, J.~Shawe-Taylor, A.~J. Smola, and R.~C.
  Williamson.
\newblock Estimating the {S}upport of a {H}igh-{D}imensional {D}istribution.
\newblock {\em Neural {C}omput.}, 13:1443--1471, 2001.

\bibitem{Scholkopf2002Learning}
B.~Sch{\"o}lkopf and A.~J. Smola.
\newblock {\em Learning with {K}ernels: {S}upport {V}ector {M}achines,
  {R}egularization, {O}ptimization, and {B}eyond}.
\newblock MIT Press, Cambridge, MA, 2002.

\bibitem{Schoelkopf2004Kernel}
B.~Sch{\"o}lkopf, K.~Tsuda, and J.-P. Vert.
\newblock {\em Kernel {M}ethods in {C}omputational {B}iology}.
\newblock MIT Press, The MIT Press, Cambridge, Massachussetts, 2004.

\bibitem{Spellman1998Comprehensive}
P.T. Spellman, G.~Sherlock, M.Q. Zhang, V.R. Iyer, K.~Anders, M.B. Eisen, P.O.
  Brown, D.~Botstein, and B.~Futcher.
\newblock Comprehensive {I}dentification of {C}ell {C}ycle-regulated {G}enes of
  the {Y}east {S}accharomyces cerevisiae by {M}icroarray {H}ybridization.
\newblock {\em Mol. {B}iol. {C}ell}, 9:3273--3297, 1998.

\bibitem{Tavazoie1999Systematic}
S.~Tavazoie, J.~D. Hughes, M.~J. Campbell, R.~J. Cho, and G.~M. Church.
\newblock Systematic determination of genetic network architecture.
\newblock {\em Nat. {G}enet.}, 1999.

\bibitem{Tegner2003Reverse}
J.~Tegner, M.~K.~S. Yeung, J.~Hasty, and J.~J. Collins.
\newblock {R}everse engineering gene networks: integrating genetic
  perturbations with dynamical modeling.
\newblock {\em Proc. Natl. Acad. Sci. USA}, 100(10):5944--5949, May 2003.

\bibitem{Tsuda2003em}
K.~Tsuda, S.~Akaho, and K.~Asai.
\newblock The em {A}lgorithm for {K}ernel {M}atrix {C}ompletion with
  {A}uxiliary {D}ata.
\newblock {\em J. {M}ach. {L}earn. {R}es.}, 4:67--81, 2003.

\bibitem{Uetz2000comprehensive}
P.~Uetz, L.~Giot, G.~Cagney, T.~A. Mansfield, R.~S. Judson, J.~R. Knight,
  D.~Lockshon, V.~Narayan, M.~Srinivasan, P.~Pochart, A.~Qureshi-Emili, Y.~Li,
  B.~Godwin, D.~Conover, T.~Kalbfleish, G.~Vijayadamodar, M.~Yang, M.~Johnston,
  S.~Fields, and J.~M. Rothberg.
\newblock A comprehensive analysis of protein-protein interactions in
  {S}accharomyces cerevisiae.
\newblock {\em Nature}, 403:623--627, 2000.

\bibitem{Vapnik1998Statistical}
V.~N. Vapnik.
\newblock {\em Statistical {L}earning {T}heory}.
\newblock Wiley, New-York, 1998.

\bibitem{Vert2007Kernel}
J.-P. Vert.
\newblock Kernel methods in genomics and computational biology.
\newblock In G.~Camps-Valls, J.-L. Rojo-Alvarez, and M.~Martinez-Ramon,
  editors, {\em Kernel Methods in Bioengineering, Signal and Image Processing}.
  IDEA Group, 2007.

\bibitem{Vert2007new}
J.-P. Vert, J.~Qiu, and W.~S. Noble.
\newblock {A} new pairwise kernel for biological network inference with support
  vector machines.
\newblock {\em BMC Bioinformatics}, 8 Suppl 10:S8, 2007.

\bibitem{Vert2005Supervised}
J.-P. Vert and Y.~Yamanishi.
\newblock Supervised graph inference.
\newblock In L.~K. Saul, Y.~Weiss, and L.~Bottou, editors, {\em Adv. {N}eural
  {I}nform. {P}rocess. {S}yst.}, volume~17, pages 1433--1440. MIT Press,
  Cambridge, MA, 2005.

\bibitem{Mering2002Comparative}
C.~von Mering, R.~Krause, B.~Snel, M.~Cornell, S.~G. Oliver, S.~Fields, and
  P.~Bork.
\newblock {C}omparative assessment of large-scale data sets of protein-protein
  interactions.
\newblock {\em Nature}, 417(6887):399--403, May 2002.

\bibitem{Yamanishi2004Protein}
Y.~Yamanishi, J.-P. Vert, and M.~Kanehisa.
\newblock Protein network inference from multiple genomic data: a supervised
  approach.
\newblock {\em Bioinformatics}, 20:i363--i370, 2004.

\bibitem{Yamanishi2005Supervised}
Y.~Yamanishi, J.-P. Vert, and M.~Kanehisa.
\newblock Supervised enzyme network inference from the integration of genomic
  data and chemical information.
\newblock {\em Bioinformatics}, 21:i468--i477, 2005.

\end{thebibliography}

\end{document}